\documentclass[sigconf,bookmarks=false]{acmart}

\AtBeginDocument{%
\providecommand\BibTeX{{%
 \normalfont B\kern-0.5em{\scshape i\kern-0.25em b}\kern-0.8em\TeX}}}

\copyrightyear{2020}
\acmYear{2020}
\setcopyright{acmcopyright}
\acmConference[ICSE-SEIP '20]{Software Engineering in Practice}{May 23--29, 2020}{Seoul, Republic of Korea}
\acmBooktitle{Software Engineering in Practice (ICSE-SEIP '20), May 23--29, 2020, Seoul, Republic of Korea}
\acmPrice{15.00}
\acmDOI{10.1145/3377813.3381367}
\acmISBN{978-1-4503-7123-0/20/05}

\usepackage{colortbl}
\usepackage{tcolorbox}

\newcommand{\bi}{\begin{itemize}}
\newcommand{\ei}{\end{itemize}}

\usepackage[tikz]{bclogo}
\newenvironment{RQ}[1]%
{\noindent\begin{minipage}[c]{\linewidth}%
\begin{bclogo}[couleur=gray!10,%
    arrondi=0.1,%
    logo=\bctrombone,%
    ombre=true]{{\small ~#1}}}%
{\end{bclogo}\end{minipage}}

\definecolor{small}{HTML}{BDC3C7}
\definecolor{medium}{HTML}{5DADE2}
\definecolor{large}{HTML}{F1C40F}
\definecolor{cuscolor}{HTML}{82E0AA}

\newcommand{\tbl}[1]{Table~\ref{tbl:#1}}
\newcommand{\fig}[1]{Figure~\ref{fig:#1}}
\definecolor{Gray}{rgb}{0.88,1,1}
\definecolor{Gray}{gray}{0.85}
\definecolor{lightgray}{gray}{0.8}

\usepackage{amsmath}

\tikzstyle{thmbox} = [rectangle, rounded corners, draw=black,
fill=Gray!20, drop shadow={fill=black, opacity=1}]

\newcommand{\quart}[4]{\begin{picture} (100,4)
 {\color{black}\put (#3,2){\circle*{4}}\put (#1,2){\line (1,0){#2}}}\end{picture}}

\newcommand{\tion}[1]{\S\ref{tion:#1}}

\begin{document}

\title{Assessing Practitioner Beliefs about Software Defect Prediction}


\author{N.C. Shrikanth, Tim Menzies}
\email{snaraya7@ncsu.edu, timm@ieee.org}
\orcid{0000-0002-1220-8533}
\affiliation{%
\institution{North Carolina State University}
\city{Raleigh}
\country{USA}
}


\begin{abstract}
Just because software developers say they believe in ``X'', that does not necessarily mean that ``X'' is true.
As shown here, there exist numerous beliefs listed in the recent Software Engineering literature which are only supported by small portions of the available data. 
Hence we ask what is the source of this disconnect between beliefs and evidence?. 

To answer this question we look for evidence for
ten beliefs within 300,000+ changes
seen in dozens of open-source projects. Some of those beliefs had strong support across all the projects; specifically,
``\textit{A commit that involves more added and removed lines is more bug-prone}'' and ``\textit{Files with fewer lines contributed by their owners  (who contribute most changes) are bug-prone}''.

Most of the widely-held beliefs studied are only sporadically supported in the data; i.e. 
large effects can appear in project data and then disappear in subsequent releases. 
Such sporadic support explains why developers believe things
that were relevant to their prior work, but not necessarily
their current work.

Our conclusion will be that we need to change the nature of the debate with Software Engineering.
Specifically, while it is important to report the effects that hold right now,
it is also important to report on what effects change over time.
\end{abstract}

 

\maketitle

\section{Introduction}

\label{sect:introduction}
Just because software developers 
say they believe in ``X'',
that does not necessarily mean that ``X'' is true.
J{\o}rgensen \& Gruschke~\cite{jorgensen09}
note that in Software Engineering (SE) domain, seldom   lessons from past projects are used to improve future reasoning  (to the detriment of new projects). 
Passos et al. note that developers often assume that lessons learned from a few past projects are general to all future projects~\cite{passos11}. 
Devanbu et al. record
opinions about software development
from 564 Microsoft software developers from around the world~\cite{devanbu2016belief}. They comment 
that programmer beliefs can  (a)~vary with each project and  (b)~may not necessarily
correspond with actual evidence in their current projects. Menzies and Nagappan et al. offer specific examples of this effect.
Nagappan et al.~\cite{nagappan2015empirical} have shown that the much-feared \textit{goto} statement is usually benign, and sometimes even useful.
Menzies et al. recently found no evidence for the \textit{delayed issue effect}~\cite{menzies2017delayed}
 (\textit{``The longer a bug remains in the system, the
exponentially more costly it becomes to fix''}) in 171 industry projects. That study  shows  that a widely held belief by both practitioners and academics   cannot be assumed to always hold.

\begin{table}[h]

\caption{ Practitioners' agreement \% on defect prediction metrics  (beliefs) 
reported in a recent IEEE Transactions on Software Engineering (TSE) paper by Wan et al.~\cite{wan2018perceptions}. 
}
\small{
\begin{tabular}{|l|p{7.5cm}|r|}
\hline
\rowcolor[HTML]{C0C0C0} 
\textbf{\#} & \multicolumn{1}{c|}{\cellcolor[HTML]{C0C0C0}\textbf{Belief}}           & \textbf{\%}  \\ \hline
\textbf{B1} & A file with a complex code change process tends to be buggy.           & 76       \\ \hline
\textbf{B2} & A file that is changed by more developers is more bug-prone.           & 64       \\ \hline
\textbf{B3} & A file with more added lines is more bug-prone.              & 61       \\ \hline
\textbf{B4} & Recently changed files tend to be buggy.                & 58       \\ \hline
\textbf{B5} & A commit that involves more added and removed lines is more bug-prone.        & 57       \\ \hline
\rowcolor[HTML]{FFFFFF} 
\textbf{B6} & Recently bug-fixed files tend to be buggy               & 49 \\ \hline
\rowcolor[HTML]{FFFFFF} 
\textbf{B7} & A file with more fixed bugs tends to be more bugprone.            & 48 \\ \hline
\rowcolor[HTML]{FFFFFF} 
\textbf{B8} & A file with more commits is more bug-prone.               & 46 \\ \hline
\rowcolor[HTML]{FFFFFF} 
\textbf{B9} & A file with more removed lines is more bug-prone.             & 35 \\ \hline
\rowcolor[HTML]{FFFFFF} 
\textbf{B10} & Files with fewer lines contributed by their owners  (who contribute most changes) 
are 
bug-prone.
& 30 \\ \hline
\end{tabular}}

\label{tbl:beliefs}
\end{table}

Accordingly, we think it is important
to always carefully evaluate developer beliefs.
For example, 
\tbl{beliefs} lists ten beliefs and software defects
listed in a recent TSE'18 paper by Wan et al.~\cite{wan2018perceptions}.
That study collected 395 responses from practitioners
to document developer beliefs about willingness to adopt technologies, challenges, defect prediction metrics, etc.
This paper looks for evidence for the \tbl{beliefs} beliefs
in 300,000+ changes seen in dozens of open-source projects.
What we find is:
\begin{itemize}
\item Two of these beliefs are supported in the data; specifically,
``\textit{A commit that involves more added and removed lines is more bug-prone}'' and ``\textit{Files with fewer lines contributed by their owners  (who contribute most changes) are bug-prone}''. 
\item Some of the beliefs supported by the data are not usually endorsed by developers.
The last column of \tbl{beliefs} shows the percent of the 395 developers who endorsed each belief. Note that the last belief  (B10) is only endorsed
by 30\% of the surveyed developers. Yet in our data, it is one of the strongest effects.
\item As to the other eight items listed in
\tbl{beliefs}, while these have
sporadic support in different releases
of the same project, there is little overall evidence for these beliefs.
\end{itemize}
Our conclusion will be that we need to change the nature of the debate with SE.
While it is important to report the effects that hold right now,
it is also important to report on what effects change over time.

This paper reports strongly supported beliefs and discusses prevalence of beliefs.  To make that argument we will use the list of developer beliefs about defect prediction in \tbl{beliefs}. 
Using that list, we will ask:

\begin{itemize}
\item[]
\textbf{RQ1 : What beliefs are strongly supported in the data?}
\begin{RQ}{Result:} We found that beliefs labeled B10 \& B5 in ~\tbl{beliefs}, which is believed by (30\% \& 57\%) of practitioners has strong support  (i.e., large correlations), whereas the more popular belief B1, which is believed by 76\% of practitioners, has relatively weaker support. 
\end{RQ}
\end{itemize}

And to identify the source of disconnect between beliefs and evidence, next we ask:
 
\begin{itemize}

\item[] \textbf{RQ2 : Does the same evidence appear
everywhere?}
\begin{itemize}
 \item Do projects show evidence for all the beliefs?
 \item Does the size of a release affect belief support?
 \item Do beliefs evolve as a project matures (more releases)?
\end{itemize}
\begin{RQ}{Result:} No, the same evidence does not appear everywhere. Only 24\% of the projects show support for all the 10 beliefs. And those projects showed support for beliefs among 10\% - 36\% of their releases. Beliefs appear stronger in smaller than larger releases, with fluctuations like B4  (weaker in \textit{large} releases but stronger in \textit{small} releases). Support for beliefs tended to decay than strengthen as the project matured.
\end{RQ}

\end{itemize}

The rest of this paper is structured as follows. \tion{related} relates the current work to the prior literature on metrics and instability. In \tion{methodology}, we discuss the choice of our datasets, it's distribution and brief about our statistical methods we use. \tion{assessing} details the construction of experiments mapping to literature. Results in \tion{results} and directions to future work in \tion{future}. \tion{threats} discusses the reliability of our findings. Finally, we summarize and discuss few implications for practice in  \tion{conclusion} and \tion{implications}.

\section{Background}\label{tion:related}

\subsection{Why Study Practitioner Beliefs?}

If we do not understand what factors lead to software defects,
then that has detrimental effects for quality assurance,
 trust, insight, training, and tool
  development.
\bi
\item {\em Quality assurance:} If we do not know what causes defects in a project, we cannot prevent those problems.
\item
 {\em Training:} Another concern is what do we train novice software engineers
  or newcomers to a project? 
  If our models are not stable, then it is hard to teach what factors most influence software quality.
\item
{\em Tool development:} Further to the last point---
if we are unsure what 
  factors most influence quality, it is difficult to design, implement and deploy tools that can successfully improve that quality (e.g. Static analysis tools).
\ei
A premise of much research~\cite{lo2015practitioners,xia2017developers,xia2019practitioners,zou2018practitioners,wan2018perceptions} is that knowledge of what factors influence software quality can be discovered by asking
software practitioners. 
But based on the results of this paper, we would say that:
\bi
\item
While such surveys produce a long list of possible effects,
\item
Only some of which might be relevant to a particular project
at a particular time.
\ei

\subsection{Related Work}

Perhaps the closest work to our work is the
work by McIntosh \& Kamei who,
in 2017,
reported that the effect on defects of code change properties based on size, entropy, history, and code ownership  changes
as software systems evolve~\cite{mcintosh2017fix}. Further, they report that although size-related metrics  (B3:\textit{Added lines} \& B9:\textit{Removed lines}) were their top contributors over other metrics, that effect
tended to fluctuate across time
 (and the strength of that effect was project-specific).

To start the research of this paper, we found five papers~\cite{lo2015practitioners,xia2017developers,xia2019practitioners,zou2018practitioners,wan2018perceptions} which list practitioner beliefs,
but do not test those beliefs with respect to empirical evidence. From this, we chose a recent qualitative study from IEEE TSE
 (see \tbl{beliefs}) paper by Wan et al. ~\cite{wan2018perceptions}.
This paper performed a comprehensive study by gathering practitioners' opinions  (395 responses) on various defect prediction research hypotheses discussed in the literature between 2012-2017. More importantly, they highlight practitioners' agreement \% on defect prediction metrics, prioritization strategy, etc. In support of that work, we work
that many of the beliefs reported in \cite{wan2018perceptions} by Wan et al. were also reported
previously in ~\cite{d2010extensive,kamei2012large}. In this paper, we revisit these beliefs independently to look for current evidence and reasoning disconnect between beliefs and evidence using the prevalence of their support.

\section{Methodology Overview}\label{tion:methodology}

In this section, we elucidate source, distribution and various attributes we collect from our sample projects. Then we detail the underlying statistical techniques we use to assess our beliefs.

\subsection{Data Source \& Cleaning}

For this work, we wanted to study conclusion instability using a much more detailed approach than the {\em Related Work} mentioned above. We analyze 3 times more changes (commits) than recent defect prediction work ~\cite{mcintosh2017fix, hoang2019deepjit} and the volume of our dataset is 8 times larger as we expand those changes (commits) that results in 301,627 source code file entries filtered from 524,851 in total.

Turning then to Github\footnote{https://github.com}, we chose 50 GitHub repository links from a recent defect prediction paper~\cite{cite_pareto} that uses Munaiah et al. GitHub project database ~\cite{munaiah2017curating}. Their work simplified choosing substantive Open Source (OS) samples from GitHub attributing to several metrics such as maturity, active developers, license, etc., We mined file-level commit histories and release information using GitHub API. To minimize bias in our modelling we applied the following sanity checks and discarded projects that have less than,

\begin{itemize}
 \item 1000 commits
 \item 10\% Bug Fixes
 \item 5 releases
 \item 30 developers
 \item 3 years of activity
\end{itemize}
And this resulted in 37 projects shown in ~\tbl{subject-systems}. Our selected projects use many current languages such as Python, Ruby, Java, JavaScript, PHP, etc. To remove noise we ignore test cases, configuration files and static resources such as text, readme, images, etc. To achieve this first we identified source code extensions from our samples they are {\em py, java, rb, c, cpp, h, php, sh, cs, scss, html,}
{\em scala, js, css, clj, ctp, erb, go, haml, hs and sql }. Essentially we consider only the file paths the ends with these extensions, with an additional check that the file/path names do not contain ``test'', to eliminate test cases.

\begin{table}
\caption{37 OS projects developed in popular programming languages. Some projects are developed using multiple programming languages.  (Rb - Ruby, Py - Python , JS - JavaScript and SCSS - Sassy-CSS)}
{\scriptsize
\begin{tabular}{|l|l|l|l|}
\hline
\rowcolor[HTML]{C0C0C0} 
\textbf{URL  (github.com/)} & \textbf{Language} & \multicolumn{1}{l|}{\cellcolor[HTML]{C0C0C0}\textbf{URL  (github.com/)}} & \multicolumn{1}{l|}{\cellcolor[HTML]{C0C0C0}\textbf{Language}} \\ \hline
\textit{activemerchant/active\_merchant} & Rb & \multicolumn{1}{l|}{\textit{xetorthio/jedis}} & \multicolumn{1}{l|}{Java,HTML} \\ \hline
\textit{activeadmin/activeadmin} & Rb,SCSS & \multicolumn{1}{l|}{\textit{spotify/luigi}} & \multicolumn{1}{l|}{Py,JS} \\ \hline
\textit{puppetlabs/beaker} & Rb,Shell & \multicolumn{1}{l|}{\textit{lra/mackup}} & \multicolumn{1}{l|}{Py} \\ \hline
\textit{boto/boto3} & Py & \multicolumn{1}{l|}{\textit{mikel/mail}} & \multicolumn{1}{l|}{Rb} \\ \hline
\textit{thoughtbot/bourbon} & SCSS,HTML & \multicolumn{1}{l|}{\textit{Seldaek/monolog}} & \multicolumn{1}{l|}{PHP} \\ \hline
\textit{bundler/bundler} & Rb,HTML & \multicolumn{1}{l|}{\textit{omniauth/omniauth}} & \multicolumn{1}{l|}{Rb,HTML} \\ \hline
\textit{teamcapybara/capybara} & Rb & \multicolumn{1}{l|}{\textit{aws/opsworks-cookbooks}} & \multicolumn{1}{l|}{Rb} \\ \hline
\textit{Codeception/Codeception} & PHP,HTML & \multicolumn{1}{l|}{\textit{thoughtbot/paperclip}} & \multicolumn{1}{l|}{Rb} \\ \hline
\textit{ooici/coi-services} & Py,C/C++ & \multicolumn{1}{l|}{\textit{getpelican/pelican}} & \multicolumn{1}{l|}{HTML,Py} \\ \hline
\textit{pentaho/data-access} & Java,JS & \multicolumn{1}{l|}{\textit{cakephp/phinx}} & \multicolumn{1}{l|}{PHP} \\ \hline
\textit{pennersr/django-allauth} & Py,HTML & \multicolumn{1}{l|}{\textit{propelorm/Propel2}} & \multicolumn{1}{l|}{PHP,HTML} \\ \hline
\textit{encode/django-rest-framework} & Py,HTML & \multicolumn{1}{l|}{\textit{puppetlabs/puppetlabs-apache}} & \multicolumn{1}{l|}{Rb} \\ \hline
\textit{django-tastypie/django-tastypie} & Py & \multicolumn{1}{l|}{\textit{sferik/rails\_admin}} & \multicolumn{1}{l|}{JS,Rb} \\ \hline
\textit{doorkeeper-gem/doorkeeper} & Rb & \multicolumn{1}{l|}{\textit{reactjs/react-rails}} & \multicolumn{1}{l|}{JS,Rb} \\ \hline
\textit{drapergem/draper} & Rb,HTML & \multicolumn{1}{l|}{\textit{resque/resque}} & \multicolumn{1}{l|}{Rb} \\ \hline
\textit{errbit/errbit} & Rb,HTML & \multicolumn{1}{l|}{\textit{restsharp/RestSharp}} & \multicolumn{1}{l|}{C\#} \\ \hline
\textit{jordansissel/fpm} & Rb & \multicolumn{1}{l|}{\textit{mperham/sidekiq}} & \multicolumn{1}{l|}{Rb,JS} \\ \hline
\textit{ros-simulation/gazebo\_ros\_pkgs} & C\&C++ & \multicolumn{1}{l|}{\textit{plataformatec/simple\_form}} & \multicolumn{1}{l|}{Rb} \\ \hline
\textit{ruby-grape/grape} & Rb,HTML \\ \cline{1-2}
\end{tabular}
}

\label{tbl:subject-systems}
\end{table}

\subsection{Data distribution}
\begin{figure}
\centering
\includegraphics[width=2.75in,keepaspectratio]{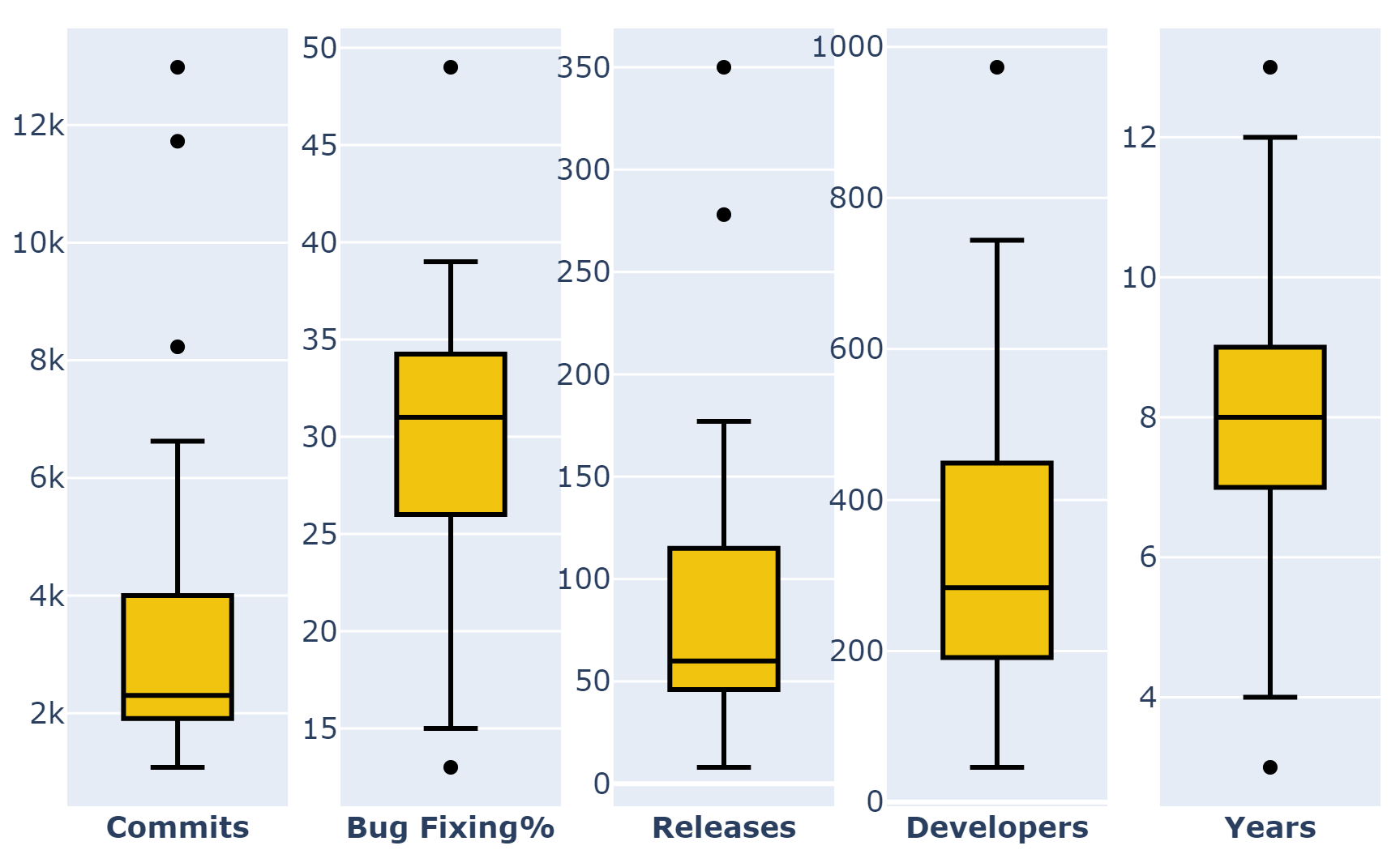}
\caption{
Distribution of Commits (2304), Bug Fixing (31\%) Releases (60), Developers (284) and Years (8) active among 37 projects. \textit{$\tilde{x}$  (median)}}\label{fig:subject_systems_distribution}
\end{figure}

Overall this sample contains 
data modified in the period 2005 to 2019 by 12,361 developers in
over 524,000 file entries. These modifications were made to 
127,950 active branch commits  (in all,
19,617,483 line insertions and 13,642,341 line deletions). On average our chosen set of projects has been active for over 7 years  (see the distributions of \fig{subject_systems_distribution}). The OS projects we chose for this study are publicly available. All our data is on-line so that other researchers can benefit  (see
\href{https://github.com/ai-se/defect\_perceptions}{\textit{github.com/ai-se/defect\_perceptions}}).

\subsection{Data Attributes}
We mine the entire commit history of a project. A tuple of our main dataset contains the following attributes unique \textit{commit\_id}, \textit{commit\_time}, the author who pushed the changes \textit{commit\_author}, affected \textit{file\_path}, number of lines inserted and deleted \textit{ (insertions, deletions)}. Importantly for every commit, we set a label Bug Fixing Commit (\textit{BFC}) as 1, if it was pushed in an attempt to fix bug (s) or 0 if otherwise. We achieve this by scanning the commit message, if they contained any derivatives of the following stemmed words like {\em bug, fix, issu, error, correct, proper, deprecat, broke, optimize, patch, solve, slow, obsolete, vulnerab, debug, perf, memory, minor, wart, better, complex, break, investigat, compile, defect, inconsist, crash, problem} or {\em resol}. We also identify releases of each project using Git releases/tags.

\subsection{Method}
Our work requires us to \textit{measure} the following two items:

\begin{itemize}
 \item \textit{Effect:} Significant associations between beliefs and defect proneness.
 \item \textit{Rank:} Significant differences among these associations.
\end{itemize}

\subsubsection{Effect:}

We use Spearman's rank correlation (a non-parametric test) to assess associations between independent variable ``$F_{BX}$" and dependent variable``$F_{D}$"  (detailed in \tion{assessing}). We chose Spearman like a previous defect prediction study ~\cite{d2010extensive} recommended to handle skewed data, further it is unaffected by transformations  (log, square-root etc) on variables. 

The Spearman's rank correlation,
\mbox{$\rho = \frac{\text{cov} (X,Y)}{\sigma_x \sigma_y}$}
between two samples $X,Y$ 
 (with means $\overline{x}$
and $\overline{y}$), as estimated using $x_i \in X$ and $y_i \in Y$ via
\[\rho = \frac{{}\sum_{i=1}^{n}  (x_i - \overline{x}) (y_i - \overline{y})}{\sqrt{\sum_{i=1}^{n}  (x_i - \overline{x})^2 (y_i - \overline{y})^2}}\]

We conclude using both correlation coefficient  ($\rho$) and its associated \textit{p\_value} in all our experiments. $\rho$ varies from 1, i.e., ranks are identical, to $-1$, i.e., ranks are the opposite, where 0 indicates no correlation.

\begin{itemize}
 \item Higher $\rho$ value indicates more evidence for defect prone.
 \item Lower $p\_value$ indicates significant result.
\end{itemize}

\subsubsection{Rank:}\label{sk}
Populations may have the same median while their distribution could be very different. Hence to identify significant differences or rank among two or more populations we use Scott-Knott test recommended by Mittas et al. in TSE'13 paper~\cite{Mittas13}. Scott-Knott is a top-down bi-clustering approach used to rank different treatments, the treatment could be beliefs, project, release or any conjunction. This method sorts a list of $l$ treatments with $\mathit{ls}$ measurements by their median score. It then splits $l$ into sub-lists \textit{m, n} in order to maximize the expected value of differences in the observed performances before and after divisions.

For lists $l,m,n$ of size $\mathit{ls},\mathit{ms},\mathit{ns}$ where $l=m\cup n$, the ``best'' division maximizes $E (\Delta)$; i.e.
the difference in the expected mean value
before and after the spit: 
\[E (\Delta)=\frac{ms}{ls}abs (m.\mu - l.\mu)^2 + \frac{ns}{ls}abs (n.\mu - l.\mu)^2\]

Further, to avoid ``small effects" with statistically significant results we employ the conjunction of bootstrapping and A12 effect size test by Vargha and Delaney ~\cite{vargha2000critique} for the hypothesis test H to check if \textit{m, n} are truly significantly different. These techniques do not make gaussian assumptions  (non-parametric).

\subsection{Terminologies}
In this section, we briefly introduce common labels, definitions, and thresholds that we weave while discussing the results.

We label beliefs as,
\begin{itemize}
 \item \textit{popular} if they have more than 50\% agreement in \tbl{beliefs}
 \item \textit{unpopular} otherwise.
\end{itemize}

\subsubsection{Evidence:} We use Spearman's rank correlation  (detailed in \tion{methodology}) $\rho$ ranges shown below to discuss strength of the evidence. From the usage of Spearman's $\rho$ in this defect prediction literature ~\cite{zimmermann2007predicting} we derive the following ranges for $|\rho|$ :
\begin{tcolorbox}
\begin{itemize}
 \item[$\ast$] 0.0 to 0.39 as \textit{no support}
 \item[$\ast$] 0.4 to 0.49 as \textit{minimum/weak support}
 \item[$\ast$] 0.5 to 0.59 as \textit{support}
 \item[$\ast$] 0.6 to 0.69 as \textit{strong support}
 \item[$\ast$] 0.7 to 1.00 as \textit{very strong support}
\end{itemize}
\end{tcolorbox}
While these ranges are debatable, we note that the results shown below are of such a ``large effect size'' that the details of these thresholds do not threaten the validity of our conclusion. But importantly these ranges have lesser impact on our conclusion as we are keen on relative ranks obtained from Scott-Knott-test of \S\ref{sk} deduced from overall population of $\rho$ scores.

\subsubsection{Project Support Population  ($P_{BX}$)}
$P_{BX}$ is the population of significant $\rho$ scores computed for a belief $BX$ beliefs 1 to 10 ie, $X\in [1..10]$ for all releases $R$ in project $P$,
Thus,
\[
 P_{BX} = \{\rho (\{F_{BX}\},\{F_{D}\})\} 
\] 
To recollect, $\rho$ is the association between:
\bi
\item $F_{BX}$, which is the metric we collect from the Github data
\item and $F_{D}$ which is the defects fixed during 6 months after the release $r$  (i.e. the defects that anyone thought were worth acting on).
\ei
We only consider significant (99\% confidence level) $\rho$ scores. This implies that some releases are not considered in the population. 

In the following, $|P_{BX}|$ is the number of releases in the project $P$, such that $|P_{BX}| \le R$. where $R$ is the total number of releases in a project $P$.

\section{Assessing Beliefs} \label{tion:assessing}
In this section, we describe the construction of our correlation experiments using the underlying rationale and relevant literature attached to each of the beliefs in ~\tbl{beliefs}. Then, we discuss the role of relevant attributes scraped from our sample projects to be used as an independent variable to compute associations.

\begin{RQ}{Objective:}
To export $P_{BX}$ which is a population of significant $\rho$ scores for a belief $BX$, where each score is computed as a correlation between, $F_{BX} \& F_{D}$ and collected over all releases $R$ in a project $P$. This is repeated for all 37 projects and 10 beliefs.
\end{RQ}

To support this objective:
\begin{itemize}
 \item[] $r:$ is a release in project $p$. We ignore first release of all projects $r>1$.
 \item[] $F:$ is a file created or modified during the pre-release $r$ period.
 \item[] $BX:$ denotes metric associated with beliefs 1 to 10, $X\in [1..10]$.
 \item[] $F_{BX}:$ denotes metric captured on file $F$
 \item[] $D:$ defects fixed 6 months after $r$  (post-release bugs)
 \item[] $F_{D}:$ defects fixed 6 months after $r$  (post-release bugs) on $F$
 
\end{itemize}

We employ the traditional release based approach to assess our beliefs. We capture metrics in the pre-release period to find associations with post-release bugs 6 months after the release. We assess the beliefs ``$BX$" for all releases in a project. We then build a population with the effect scores collected in all releases and projects to be analyzed in the \tion{results}. In each release, we compute the belief metric for each distinct file ``$F$'' modified in a release $r$ where $r \in R$ (total number of releases). And defect proneness ``$D$'' is the number of post-release bugs. Then ``$F_{D}$'' is the number of post-release  ($r+6$months) bugs on file ``$F$''. Hence, ``$F_{D}$'' computation is common across all the beliefs whereas ``$F_{BX}$''  (belief metric on a file) captured during the pre-release period $r$ is elucidated in the following sections.

In the following, we discuss the specifics of how to assess
the beliefs
of \tbl{beliefs}.

\subsection{B1: Complex Code Changes} 

\textbf{Description:} ``\textit{A file with a complex code change process tends to be buggy}". In 2009, Hassan \cite{entropy} adopted Shannon Entropy from information theory to devise few code change models that weigh scattered modifications (complexity) of a file using its change history. The intuition behind this belief is that change scattering of a file over some periods makes it cumbersome for developers to maintain, thus making it defect prone. We use Hassan's compute History Complexity Metric  ($HCM^{1d}$) with the decay factor  ($d$) to assess this belief. The decay factor is used to undermine earlier modifications. 

\textbf{Procedure:} We divide the pre-release period $r$ into bi-weekly (14 days) periods to compute entropy $HCM$ for a file $F$ and this will be $F_{B1}$. In cases where releases have less than 14 days, $r$ will have just two equal size periods. We then export the correlation between $F_{B1}$ \& $F_{D}$.

\subsection{B2 \& B10: Ownership}
\textbf{Description:} B2:``\textit{Files changed by more developers are more buggy}''. In 2010, Matsumoto et al. ~\cite{matsumoto2010analysis} showed human factors can be used to forecast defects. Specifically, they saw that files touched by more developers make the file prone to defects.

B10:``\textit{Files with fewer lines contributed by their owners  (who contribute most changes) are bug-prone}''. Bird et al. in their work ~\cite{bird2011don} explore various ownership related metrics. We employ their idea of minor \& major contributor, where many minor contributors modifying a file makes it defect prone. They define a minor contributor as someone who has made less than 5\% changes and a major contributor as someone who has made at-least 5\% or more. Thus if a file in a release has changed by a large proportion of minor contributors it is prone to more defects.

\textbf{Procedure:} $F_{B2}$ here is the count of distinct \textit{commit authors} who made some change to the file in the pre-release. $F_{B10}$ is the \% of minor contributors  (who contributed less than 5\% code churn) for file $F$ in the pre-release. 
We then export the correlation between $F_{B2}$ \& $F_{D}$ and $F_{B10}$ \& $F_{D}$.

\subsection{B3 \& B9 : \textit{Code Churn}}
\textbf{Description:} B3:``\textit{A file with more added lines is more bug-prone}'' and B9:``\textit{A file with more removed lines is more bug-prone}''. In 2005 Nagappan \& Ball in ~\cite{nagappan2005use} showed that relative code churns of files using number of added or deleted lines between subsequent versions are good indicators to forecast defects. 

\textbf{Procedure:} We measure $F_{B3}$ or $F_{B9}$ for a file $F$ by aggregating on the \textit{number of lines added  (B3)} or \textit{number of lines deleted  (B9)} only during the pre-release period $r$. We then export the correlation between $F_{B3}$ \& $F_{D}$ and $F_{B9}$ \& $F_{D}$.

\subsection{B4 \& B6: \textit{Temporal}}
\textbf{Description:} Two of these temporal heuristics are introduced and explored by Hassan et al. \cite{cite_top_10_more_recent} in their popular work, ``Top 10 list for dynamic defect prediction". B4:``\textit{Recently changed files tend to be buggy}'' and B6:``\textit{Recently bug-fixed files tend to be buggy}". The rationale here is that files modified closer to release periods may not be tested effectively and as a result, recently changed files would tend to introduce more bugs  (also see ~\cite{graves2000predicting}) in the near future  (B4). Another intuition is that faults tend to arise at the same spot is a good indicator to measure defect proneness (B6).

\textbf{Procedure:} Attributes \textit{commit\_time} (longer value indicates more recent) and \textit{BFC}  (1 indicates bug fixing, 0 otherwise) are used to model these two beliefs. But a file can be modified (committed) multiple times in the pre-release period. Hence, we assign the maximum \textit{commit\_time} for $F_{B4}$, similarly we assign the maximum \textit{commit\_time} for $F_{B6}$; provided \textit{BFC = 1} in the pre-release period. Further in $F_{B6}$ if a file is never modified for the purpose of fixing a defect in the pre-release period, its ignored for analysis. We then export the correlation between $F_{B4}$ \& $F_{D}$ and $F_{B6}$ \& $F_{D}$.

\subsection{B5: Commit Churns}

\textbf{Description:} ``\textit{A commit that involves more added and removed lines is more bug-prone.}". A single commit affects one or more files with some line additions and/or deletions. Hindle et al. ~\cite{hindle2008large} and Hattori et al. ~\cite{hattori2008nature} studied the nature and distribution of large commits in terms of the number of files it changed and highlighted that although large commits are rare they are unsafe to ignore. A commit can push the same (or more) amount of line changes  with fewer files. Thus rather than the number of files a commit changes, we weigh this belief based on the amount of added and removed lines a commit pushes into the system.

\textbf{Procedure:} Unlike other beliefs, that can be measured at file-level, a commit is a collection of files. It's immutable, meaning it cannot be directly traced in the post-release period. Hence, we assess the impact of a commit using the files it affected. Thus, the independent variable is a commit in the pre-release period $F_{B5}$ (represents a commit), which is an aggregate of code churn  (insertions + deletions) for all the files part of the commit. Similarly, commit defect proneness is the aggregation of file defect proneness $F_{D}$ in the post-release period. In other words, we collectively correlate between a file $F$ that is part of a large commit in the pre-release period with the number of bugs introduced by $F$ in the post-release period. We then export the correlation between $F_{B5}$ \& $F_{D}$.

\subsection{B7 \& B8: \textit{Something More}}
\textbf{Description:} Graves et al. in ~\cite{graves2000predicting} found an effect among the two process-related metric B7 \& B8 ie., ``\textit{A file with more fixed bugs tends to be more bug-prone}'' and ``\textit{A file with more commits is more bug-prone}'' through analyzing code from a telephone switching system.

\textbf{Procedure:} $F_{B7}$ is the count of bug fixes on file $F$ and $F_{B8}$ is the count of modifications (commits) made on file $F$, computed during their corresponding pre-release periods. We then export the correlation between $F_{B7}$ \& $F_{D}$ and $F_{B8}$ \& $F_{D}$.

\section{Results}\label{tion:results}

In this section, we explore two of our RQ's. For each RQ we discuss the motivation, approach, and findings.

\subsection{RQ1: What beliefs are strongly supported in the data?}

\textbf{Motivation:} Similar to the work by Devanbu et al. in ~\cite{devanbu2016belief} we compare practitioners' agreement \% with empirical evidence. Our objective here is to find beliefs that have strong support.

\textbf{Approach:} We compute $P_{BX}$ for each belief $BX$ for all the 37 projects. This results in 10 independent $P_{BX}$ populations one for each belief. Next, we rank these populations by their median and effect difference using the Scott-Knott-test of \S\ref{sk}. That results in \tbl{sk_overall_correlation} which we compare with practitioners' agreement \% in ~\tbl{beliefs}.

\begin{table}[h]
\centering
\caption{Scott-Knott test applied to all the 10 beliefs, that contains support $\rho$ scores of 3,198 releases in all the 37 projects. Each row represents a population of $P_{BX}$ scores across all 37 projects.  (Higher rank indicates stronger support). IQR - Interquartile Range.
}\label{tbl:sk_overall_correlation}
\begin{tabular}{|l|l|r|r|l|}
\hline
\rowcolor[HTML]{C0C0C0} 
\rowcolor[gray]{.9} \textbf{Rank} & \textbf{Belief} & \textbf{Median} & \textbf{IQR} & \\
 1 &  B9 (35\%) & 43 & 27 & \quart{28}{27}{43}{100} \\
 1 &  B4 (58\%) & 44 & 28 & \quart{31}{28}{44}{100} \\
 \hline
 2 &  B3 (61\%) & 45 & 21 & \quart{35}{21}{45}{100} \\
 2 &  B1 (76\%) & 46 & 21 & \quart{35}{21}{46}{100} \\
 2 &  B6 (49\%) & 49 & 47 & \quart{27}{47}{49}{100} \\
 2 &  B7 (48\%) & 49 & 22 & \quart{38}{22}{49}{100} \\
 2 &  B2 (64\%) & 50 & 21 & \quart{39}{21}{50}{100} \\
 2 &  B8 (46\%) & 51 & 22 & \quart{39}{22}{51}{100} \\
 \hline
 3 &  B10 (30\%) & 63 & 23 & \quart{50}{23}{63}{100} \\
 \hline
 4 &  B5 (57\%) & 71 & 26 & \quart{57}{26}{71}{100} \\\hline
\end{tabular}
\end{table}

\textbf{Findings:} Using the results in \tbl{sk_overall_correlation} we report \textit{popular} belief B10 and \textit{unpopular} belief B5 have significantly \textit{strong} support. Surprisingly, these \textit{unpopular} beliefs B6,B7,B8 and B10 have better support than some \textit{popular} beliefs B1, B3 \& B4. For example:

\bi
\item
The ownership-based belief B10 with only 30\% practitioner agreement shows \textit{strong support}.
\item
The temporal belief B4 though with 58\% practitioner agreement has relatively weaker support 0.4. B1 with the largest practitioner agreement 76\% shows weak support.
\ei

Overall, only a few beliefs B2,B5 \& B9 are in agreement (\%) with practitioners beliefs. Of those, B5:\textit{Large Commits}, with 57\% practitioner agreement has the highest effect  (0.7) among all the beliefs.

\begin{RQ}{Result:} We found that beliefs labeled B10 \& B5 in ~\tbl{beliefs}, which is believed by  (30\% \& 57\%) of practitioners has strong support  (i.e., large correlations), whereas the more popular belief B1, which is believed by 76\% of practitioners, has relatively weaker support. 
\end{RQ}

\textbf{\subsection{RQ2 : Does the same evidence appear everywhere?}}
\begin{figure}[h]
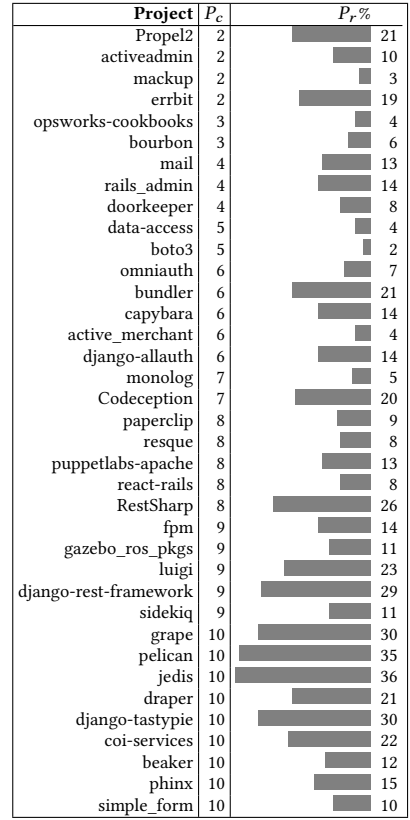

{
\footnotesize
\renewcommand{\baselinestretch}{0.7}
\frame{
\begin{tabular}{r|r|rr}
\textbf{Project} & \textbf{$P_{c}$} & \textbf{$P_{r}\%$} \\\hline
Propel2&\rotatebox{0}{2}&\textcolor{gray}{\rule{10.5mm}{2mm}}&\rotatebox{0}{21}\\
activeadmin&\rotatebox{0}{2}&\textcolor{gray}{\rule{5.0mm}{2mm}}&\rotatebox{0}{10}\\
mackup&\rotatebox{0}{2}&\textcolor{gray}{\rule{1.5mm}{2mm}}&\rotatebox{0}{3}\\
errbit&\rotatebox{0}{2}&\textcolor{gray}{\rule{9.5mm}{2mm}}&\rotatebox{0}{19}\\
opsworks-cookbooks&\rotatebox{0}{3}&\textcolor{gray}{\rule{2.0mm}{2mm}}&\rotatebox{0}{4}\\
bourbon&\rotatebox{0}{3}&\textcolor{gray}{\rule{3.0mm}{2mm}}&\rotatebox{0}{6}\\
mail&\rotatebox{0}{4}&\textcolor{gray}{\rule{6.5mm}{2mm}}&\rotatebox{0}{13}\\
rails\_admin&\rotatebox{0}{4}&\textcolor{gray}{\rule{7.0mm}{2mm}}&\rotatebox{0}{14}\\
doorkeeper&\rotatebox{0}{4}&\textcolor{gray}{\rule{4.0mm}{2mm}}&\rotatebox{0}{8}\\
data-access&\rotatebox{0}{5}&\textcolor{gray}{\rule{2.0mm}{2mm}}&\rotatebox{0}{4}\\
boto3&\rotatebox{0}{5}&\textcolor{gray}{\rule{1.0mm}{2mm}}&\rotatebox{0}{2}\\
omniauth&\rotatebox{0}{6}&\textcolor{gray}{\rule{3.5mm}{2mm}}&\rotatebox{0}{7}\\
bundler&\rotatebox{0}{6}&\textcolor{gray}{\rule{10.5mm}{2mm}}&\rotatebox{0}{21}\\
capybara&\rotatebox{0}{6}&\textcolor{gray}{\rule{7.0mm}{2mm}}&\rotatebox{0}{14}\\
active\_merchant&\rotatebox{0}{6}&\textcolor{gray}{\rule{2.0mm}{2mm}}&\rotatebox{0}{4}\\
django-allauth&\rotatebox{0}{6}&\textcolor{gray}{\rule{7.0mm}{2mm}}&\rotatebox{0}{14}\\
monolog&\rotatebox{0}{7}&\textcolor{gray}{\rule{2.5mm}{2mm}}&\rotatebox{0}{5}\\
Codeception&\rotatebox{0}{7}&\textcolor{gray}{\rule{10.0mm}{2mm}}&\rotatebox{0}{20}\\
paperclip&\rotatebox{0}{8}&\textcolor{gray}{\rule{4.5mm}{2mm}}&\rotatebox{0}{9}\\
resque&\rotatebox{0}{8}&\textcolor{gray}{\rule{4.0mm}{2mm}}&\rotatebox{0}{8}\\
puppetlabs-apache&\rotatebox{0}{8}&\textcolor{gray}{\rule{6.5mm}{2mm}}&\rotatebox{0}{13}\\
react-rails&\rotatebox{0}{8}&\textcolor{gray}{\rule{4.0mm}{2mm}}&\rotatebox{0}{8}\\
RestSharp&\rotatebox{0}{8}&\textcolor{gray}{\rule{13.0mm}{2mm}}&\rotatebox{0}{26}\\
fpm&\rotatebox{0}{9}&\textcolor{gray}{\rule{7.0mm}{2mm}}&\rotatebox{0}{14}\\
gazebo\_ros\_pkgs&\rotatebox{0}{9}&\textcolor{gray}{\rule{5.5mm}{2mm}}&\rotatebox{0}{11}\\
luigi&\rotatebox{0}{9}&\textcolor{gray}{\rule{11.5mm}{2mm}}&\rotatebox{0}{23}\\
django-rest-framework&\rotatebox{0}{9}&\textcolor{gray}{\rule{14.5mm}{2mm}}&\rotatebox{0}{29}\\
sidekiq&\rotatebox{0}{9}&\textcolor{gray}{\rule{5.5mm}{2mm}}&\rotatebox{0}{11}\\
grape&\rotatebox{0}{10}&\textcolor{gray}{\rule{15.0mm}{2mm}}&\rotatebox{0}{30}\\
pelican&\rotatebox{0}{10}&\textcolor{gray}{\rule{17.5mm}{2mm}}&\rotatebox{0}{35}\\
jedis&\rotatebox{0}{10}&\textcolor{gray}{\rule{18.0mm}{2mm}}&\rotatebox{0}{36}\\
draper&\rotatebox{0}{10}&\textcolor{gray}{\rule{10.5mm}{2mm}}&\rotatebox{0}{21}\\
django-tastypie&\rotatebox{0}{10}&\textcolor{gray}{\rule{15.0mm}{2mm}}&\rotatebox{0}{30}\\
coi-services&\rotatebox{0}{10}&\textcolor{gray}{\rule{11.0mm}{2mm}}&\rotatebox{0}{22}\\
beaker&\rotatebox{0}{10}&\textcolor{gray}{\rule{6.0mm}{2mm}}&\rotatebox{0}{12}\\
phinx&\rotatebox{0}{10}&\textcolor{gray}{\rule{7.5mm}{2mm}}&\rotatebox{0}{15}\\
simple\_form&\rotatebox{0}{10}&\textcolor{gray}{\rule{5.0mm}{2mm}}&\rotatebox{0}{10}\\
\end{tabular}
}
}
\caption{ Coverage of beliefs $P_{c}$ and its prevalence among releases $P_{r}$ is shown here. Projects are sorted based on $P_{C}$. $P_{C}$ is the count of beliefs that show at least a \textit{minimum support} in the project. $P_{r}\%$ is the proportion of aggregated experiments that show  at least a \textit{minimum support} in the project for all the beliefs. }
\label{fig:belief_project_coverage}
\end{figure}

\begin{table}[h]
\centering
\caption{
\frame{
\fcolorbox{small}{small}{}Small (S)
\fcolorbox{medium}{medium}{}Medium (M)
\fcolorbox{large}{large}{}Large (L)
}
Scott-Knott-test placed 30 treatments  (10 beliefs $\ast$ 3 release size) into various ranks using their support $P_{BX}$ population. Beliefs with high support $\rho$ are found in the bottom and less support in the top.  (Higher rank indicates stronger support). IQR - Interquartile Range. Treatment labels are sub-scripted with practitioners' agreement from \tbl{beliefs}.
}
\label{tbl:sk_belief_fluctuations}

\begin{tabular}{|l|l|r|r|l|}
\footnotesize
\cellcolor{lightgray}\textbf{Rank} & \cellcolor{lightgray}\textbf{Treatment} & \cellcolor{lightgray}\textbf{Median} & \cellcolor{lightgray}\textbf{IQR} & \cellcolor{lightgray} \\
 1 &  \cellcolor{large}$L_{B9 (35\%)}$ & 33 & 21 & \quart{21}{21}{33}{100} \\
 1 &  \cellcolor{large}$L_{B6 (49\%)}$ & 34 & 23 & \quart{23}{23}{34}{100} \\
 1 &  \cellcolor{large}$L_{B4 (58\%)}$ & 37 & 22 & \quart{26}{22}{37}{100} \\ \hline
 2 &  \cellcolor{large}$L_{B1 (76\%)}$ & 38 & 17 & \quart{31}{17}{38}{100} \\
 2 &  \cellcolor{large}$L_{B3 (61\%)}$ & 38 & 14 & \quart{31}{14}{38}{100} \\
 2 &  \cellcolor{large}$L_{B8 (46\%)}$ & 41 & 19 & \quart{33}{19}{41}{100} \\
 2 &  \cellcolor{large}$L_{B2 (64\%)}$ & 42 & 18 & \quart{32}{18}{42}{100} \\
 2 &  \cellcolor{large}$L_{B7 (48\%)}$ & 42 & 19 & \quart{33}{19}{42}{100} \\ \hline
 3 &  \cellcolor{large}$L_{B10 (30\%)}$ & 52 & 22 & \quart{40}{22}{52}{100} \\
 3 &  \cellcolor{medium}$M_{B9 (35\%)}$ & 52 & 14 & \quart{45}{14}{52}{100} \\
 3 &  \cellcolor{medium}$M_{B4 (58\%)}$ & 54 & 16 & \quart{45}{16}{54}{100} \\
 3 &  \cellcolor{medium}$M_{B7 (48\%)}$ & 54 & 14 & \quart{49}{14}{54}{100} \\
 3 &  \cellcolor{medium}$M_{B3 (61\%)}$ & 54 & 12 & \quart{49}{12}{54}{100} \\
 3 &  \cellcolor{medium}$M_{B2 (64\%)}$ & 55 & 14 & \quart{48}{14}{55}{100} \\
 3 &  \cellcolor{medium}$M_{B8 (46\%)}$ & 55 & 13 & \quart{49}{13}{55}{100} \\
 3 &  \cellcolor{medium}$M_{B1 (76\%)}$ & 55 & 17 & \quart{47}{17}{55}{100} \\ 
 3 &  \cellcolor{medium}$M_{B6 (49\%)}$ & 55 & 24 & \quart{41}{24}{55}{100} \\  \hline
 4 &  \cellcolor{large}$L_{B5 (57\%)}$ & 63 & 25 & \quart{49}{25}{63}{100} \\ \hline
 5 &  \cellcolor{medium}$M_{B10 (30\%)}$ & 65 & 15 & \quart{56}{15}{65}{100} \\
 5 &  \cellcolor{small}$S_{B1 (76\%)}$ & 66 & 4 & \quart{64}{4}{66}{100} \\
 5 &  \cellcolor{medium}$M_{B5 (57\%)}$ & 69 & 23 & \quart{57}{23}{69}{100} \\
 5 &  \cellcolor{small}$S_{B2 (64\%)}$ & 74 & 13 & \quart{69}{13}{74}{100} \\
 5 &  \cellcolor{small}$S_{B6 (49\%)}$ & 75 & 22 & \quart{64}{22}{75}{100} \\
 5 &  \cellcolor{small}$S_{B9 (35\%)}$ & 74 & 15 & \quart{69}{15}{74}{100} \\
 5 &  \cellcolor{small}$S_{B3 (61\%)}$ & 76 & 19 & \quart{69}{19}{76}{100} \\
 5 &  \cellcolor{small}$S_{B7 (48\%)}$ & 78 & 16 & \quart{70}{16}{78}{100} \\ \hline
 6 &  \cellcolor{small}$S_{B10 (30\%)}$ & 80 & 15 & \quart{72}{15}{80}{100} \\
 6 &  \cellcolor{small}$S_{B4 (58\%)}$ & 82 & 20 & \quart{70}{20}{82}{100} \\
 6 &  \cellcolor{small}$S_{B8 (46\%)}$ & 80 & 16 & \quart{71}{16}{80}{100} \\
 6 &  \cellcolor{small}$S_{B5 (57\%)}$ & 87 & 18 & \quart{77}{18}{87}{100} \\ \hline
\end{tabular}
\end{table}

\textbf{Motivation:} The above results show that, overall, many commonly held beliefs do not always hold across all the data. But this is not to say
that {\em sometimes} it may be true that {\em some} of the beliefs
of \tbl{beliefs} are not true. 

Devanbu et al. \cite{devanbu2016belief} found that practitioners had different opinions working for different projects, within the same organization. To explore the source of that disconnect, we wish to investigate irregularities of evidence among projects, releases and over time, thus we ask :

\begin{itemize}
 \item[] a) Do projects show evidence for all the beliefs?
 \item[] b) Does the size of a release affect belief support?
 \item[] c) Do beliefs evolve as a project matures (more releases)?
\end{itemize}

\textbf{a) Do projects show evidence for all the beliefs ?}

\textbf{Approach:} For each project we measure coverage of beliefs $P_{C}$, which is simply the number of beliefs a project $P$ shows at least a \textit{minimum support}. In other words a project $P$ covers a belief $BX$ if $\tilde{x} (P_{BX}) \geq 0.4$. Then, we measure $P_{r}\%$  (release prevalence) which is the proportion of our aggregated correlation experiments that shows at least a \textit{minimum support}. To compute this we aggregate $P_{BX}$ for all 10 beliefs for a project $P$. Then using this aggregated population  we compute the \% of $\rho \geq 0.4$ . 

\textbf{Findings:} The results in \fig{belief_project_coverage} show that:
\bi
\item
Only 24\% of the projects show support for all the 10 beliefs. 
\item
Only
24\% of projects
show support for less than 5 beliefs
 (but projects that covered all 10 beliefs had their evidence in only   (10 - 36\%) of its releases).
\ei
\textbf{b) Does the size of a release affect belief support? }

\textbf{Approach:} We define size of a release based on the number of files created or modified in a release. To measure fluctuations, first we group releases into three non-overlapping categories namely \textit{small, medium \& large} based on the number of distinct files $D_{F}$ in a release. Using the distribution of $D_{F}$ among releases of all the 37 projects, we find the  $\tilde{x} (D_{F})$ to be 18 (files). And using the Inter-Quartile range of this distribution, we place a release $r$ of a project $P$ in one of the following categories:
\begin{itemize}
 \item \textit{small:} If, $3 < D_{F} < 18$  ($\tilde{x}$)
 \item \textit{medium:} If, $18 \leq D_{F} < Q3$  (Third quartile)
 \item \textit{large:} If, $D_{F} \geq Q3$  (Third quartile)
\end{itemize}

Then, for each belief $BX$ we segregate the support score population $P_{BX}$  (computed in RQ1) into three different populations based on size of the release. After grouping, this results in 30 populations (treatments) which we cluster using Scott-Knott-test\S\ref{sk}, resulting in \tbl{sk_belief_fluctuations}. 
Note that we group releases based on the number of files $D_{F}$ rather than the median duration (21 days), since we model our experiments based on files $F$. This decision should not affect our experiment as we tested $D_{F}$ and its corresponding release duration to be positively correlated with \textit{strong support} of 0.6 (median); considering all releases in all the 37 projects.

\textbf{Findings:} We observe a clear drift in effect distribution among the three release sizes in \tbl{sk_belief_fluctuations}, indicating size of a release affects belief support. For example, temporal belief B4 shows strong support in \textit{small} releases but not in \textit{large} releases. Notably, smaller releases have better support than larger releases. Thus, in summary, \tbl{sk_belief_fluctuations} tells us that we can reason better about smaller releases than larger releases.

That said, it is important to add
that while  47\% of releases in our projects are \textit{small}, only 18\%  (718 releases) of them were qualified for the treatments above as the rest were insignificant  ($p\_value \geq 0.01$). So while our conclusions about small releases are a promising result, overall across all our data, it holds for a very small group.

\textbf{c) Do beliefs evolve as a project matures (more releases)?}

\textbf{Approach:} To observe whether belief support tend to strengthen or decay as a project matures with more releases,
we test if there is a linear relationship between belief support and its maturity. To measure this, we correlate between all release dates ($\{r_{date}\}$) of a project with its corresponding support scores $P_{BX}$ for a belief $BX$.

The $Growth (\%)$ for a belief is the proportion of projects that shows a correlation $\rho \geq 0.4$  (indicating a positive relationship between release dates and belief support). Similarly, $Decay (\%)$ for a belief is the proportion of projects that shows a correlation $\rho \leq -0.4$  (indicating a negative relationship between release dates and belief support).
\begin{figure*}[h]
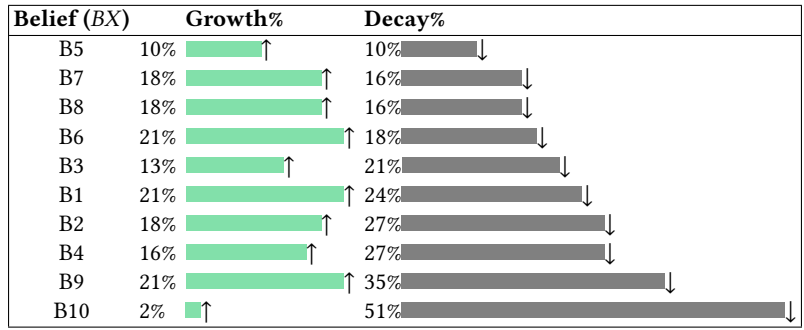

\centering


\frame{
\begin{tabular}{cllll}
\textbf{Belief ($BX$)}&&\textbf{Growth\%}& \textbf{Decay\%}\\ \hline

B5&10\% &\textcolor{cuscolor}{\rule{10mm}{2mm}}$\uparrow$&10\%\textcolor{gray}{\rule{10mm}{2mm}}$\downarrow$\\
B7&18\% &\textcolor{cuscolor}{\rule{18mm}{2mm}}$\uparrow$&16\%\textcolor{gray}{\rule{16mm}{2mm}}$\downarrow$\\
B8&18\% &\textcolor{cuscolor}{\rule{18mm}{2mm}}$\uparrow$&16\%\textcolor{gray}{\rule{16mm}{2mm}}$\downarrow$\\
B6&21\% &\textcolor{cuscolor}{\rule{21mm}{2mm}}$\uparrow$&18\%\textcolor{gray}{\rule{18mm}{2mm}}$\downarrow$\\
B3&13\% &\textcolor{cuscolor}{\rule{13mm}{2mm}}$\uparrow$&21\%\textcolor{gray}{\rule{21mm}{2mm}}$\downarrow$\\
B1&21\% &\textcolor{cuscolor}{\rule{21mm}{2mm}}$\uparrow$&24\%\textcolor{gray}{\rule{24mm}{2mm}}$\downarrow$\\
B2&18\% &\textcolor{cuscolor}{\rule{18mm}{2mm}}$\uparrow$&27\%\textcolor{gray}{\rule{27mm}{2mm}}$\downarrow$\\
B4&16\% &\textcolor{cuscolor}{\rule{16mm}{2mm}}$\uparrow$&27\%\textcolor{gray}{\rule{27mm}{2mm}}$\downarrow$\\
B9&21\% &\textcolor{cuscolor}{\rule{21mm}{2mm}}$\uparrow$&35\%\textcolor{gray}{\rule{35mm}{2mm}}$\downarrow$\\
B10&2\% &\textcolor{cuscolor}{\rule{2mm}{2mm}}$\uparrow$&51\%\textcolor{gray}{\rule{51mm}{2mm}}$\downarrow$\\

\end{tabular}}
\caption{Evolution of beliefs among all 37 projects. Growth \% is the proportion of projects where a corresponding belief correlated with $\rho \ge 0.4 $ between project release dates and empirical evidence ($P_{BX}$). And, Decay \% is the proportion of projects where a corresponding belief correlated with $\rho \le -0.4 $ between project release dates and empirical evidence ($P_{BX}$). Beliefs in the plot are sorted based on Decay \%.
\frame{
\fcolorbox{cuscolor}{cuscolor}{}$\uparrow$\textit{ Growth} \textit{ \& }
\fcolorbox{gray}{gray}{}$\downarrow$\textit{ Decay} 
}
}
\label{fig:belief_exp}
\end{figure*}

\textbf{Findings:} \fig{belief_exp} show some support for beliefs both decaying and growing. Beliefs B6 \& B9 show growth in 21\% of the projects.
Four of the beliefs  (B2, B4, B9 \& B10) are decaying in more than 25\% of the projects with the highest decay of 51\% is observed with ownership-based belief \textit{B10} and notably the growth of just 2\%.

That said, the major trend in \fig{belief_exp} is that
beliefs tend to decay, not strengthen as a project matures. Further, we highlight that we only assessed for linearity  and we don't delve into the magnitude of this effect which would remain a future work. 

\begin{RQ}{Result:} No, the same evidence does not appear everywhere. Only 24\% of the projects show support for all the 10 beliefs. And those projects showed support for beliefs among 10\% - 36\% of their releases. Beliefs appear stronger in smaller than larger releases, with fluctuations like B4  (weaker in \textit{large} releases but stronger in \textit{small} releases). Support for beliefs tended to decay than strengthen as the project matured.
\end{RQ}

\section{Future directions}\label{tion:future}
We portrayed various irregularities of beliefs in data. Krishna et al. through transfer learning techniques~\cite{krishna2016too,krishna2018bellwethers} showed instabilities can be minimized by identifying a representative oracle project among `$N$' projects. But in our opinion, as a project evolves it is more likely that such an oracle may need frequent adjustments or even replacements. Thus in a result analogous to  Menzies et al. in ~\cite{menzies2011local}, we advise focusing on factors that help to answer when \& where support for beliefs holds for our future work.

Other researchers endorse our call
for more reasoning about the context in SE. Recently Zhang et al. in ~\cite{zhang2014towards} showed improved global predictive power by including 6 context factors  ( i.e., programming language, 
issue tracking, the total lines of code, the total number of
files, the total number of commits, and the total number
of developers). Notably only 3 out of 21 defect prediction works cited Petersen \& Wohlin's context factors ~\cite{petersen2009context} (2009) in the past decade. 


\section{Threats to validity}
\label{tion:threats}


\subsection{Threats to internal validity}

This paper only explored 10 of the 15 metric-related beliefs documented by Wan et al.~\cite{wan2018perceptions} in \tbl{beliefs}.
We found that some of the modeling decisions about how to map data into \tbl{beliefs} required extensive, possibly even arcane, explanations. Objectively, we were able to answer our central question using the 10 beliefs and it would remain the same with or without exploring additional beliefs. Like this prominent~\cite{ray2014large} and a recent ~\cite{cite_pareto} large scale analysis, we rely on labeling commit messages for bug fixes as our independent variable. Due to limitations in the heuristics used to generate those labels~\cite{cite_szz_original,fan2019impact}, such
labels might be misleading. To partially mitigate this issue of false negatives, we expanded our set of keywords for better coverage, which we detailed in \tion{methodology}. To validate if false positives impacted our results, we cross-checked whether projects with higher \textit{Bug Fix \%} were likely to show support for more beliefs.

\subsection{Threats to external validity} 
Our work is biased by the projects and data found in our
sample. This means that our conclusions
could differ from other publications.
For example, many of those publications discussed
C and C++ projects while our sample contained many Ruby projects. Having said this, our sample sufficiently covers many popular programming languages like Python, Java, etc. All our projects are OS but Agrawal et al. in ~\cite{agrawal2018we} showed that many open-source lessons hold for in-house. We identified releases using \textit{git tags} which mark a boundary for readiness in the commit. Some of these changes were too small  (just a few hours) even for rapid releases ~\cite{mantyla2015rapid}.
Hence, we only consider releases that had at least 3 distinct files changed.

\subsection{Threats to statistical conclusion validity}
Our conclusions are based on correlation which means
the strength  (or weakness) of this analysis is the same as the strength or weakness of correlation.
To increase the validity of those conclusions,
we only reported significant at the 0.01 level. 
Also, we ignored correlations with less than 4 observations even at a significant level. As some releases have less than 3 files changed and technically it is possible to get high correlation with a zero \textit{p\_value} with just two observations, but we considered this unwise  (so we deleted such conclusions).

\section{Summary}
\label{tion:conclusion}

At the start of a software analytics project, it is important to focus software analytics on questions of interest to the client.
Therefore, it is very important to document developer beliefs, as done by Wan et al. ~\cite{wan2018perceptions}. That being said, once project data becomes available, it is just as important to update the focus in accordance with the observed effects. 

In this paper, we report the source of the disconnect between practitioners' perception of defect prediction metrics and empirical evidence. We argue sporadicity of evidence prevalence is the leading cause of the disconnect between stated beliefs  (e.g. the Wan et al. ~\cite{wan2018perceptions} paper) and observed evidence. Despite irregularities, we offer few beliefs that stood out B5 (Large Commits) and B10 (Owner contribution) with good prevalence in our large scale study. Thus we encourage developers to update their beliefs when the evidence demands it. This paper also observed fluctuations of evidence in a new dimension  (size of a release).

\section{Implications for Practice}
\label{tion:implications}
Inferring from the results in \tion{results} we say:
\bi
\item Yes, we should study working software engineers to learn a list of effects that might damage software quality;
\item But no, do not assume that all those effects hold at all times over the current project.
\ei
More specifically, when setting policies for software projects, managers and project leads should not ``bet'' on a small number of effects. Rather, they
should:
\bi
\item 
Perpetually monitor for the presence, or absence of a range of effects  (such as those listed in \tbl{beliefs});
\item
Perpetually adjust their code review and code refactoring processes such that they learn  (a)~not to trigger on old effects that now no longer hold or  (b)~trigger on new
effects that have just appeared.
\ei

\begin{acks}
This work was partially supported by NSF grant \#1908762. We would like to thank the anonymous reviewers for their valuable feedback.
\end{acks}

\balance
\bibliographystyle{ACM-Reference-Format}
\bibliography{references}

\end{document}